# Electroforming-Free TaOx Memristors using Focused Ion Beam Irradiations

J. L. Pacheco, D. L. Perry, D. R. Hughart, M. Marinella and E. Bielejec

Sandia National Laboratories, Albuquerque, New Mexico 87185, USA.

**Abstract:**
We demonstrate creation of electroforming-free TaOx memristive devices using focused ion beam irradiations to locally define conductive filaments in TaOx films. Electrical characterization shows that these irradiations directly create fully functional memristors without the need for electroforming. Ion beam forming of conductive filaments combined with state-of-the-art nano-patterning presents a CMOS compatible approach to wafer level fabrication of fully formed and operational memristors.

Scaling of existing Silicon based memory technologies to ultra-dense applications presents a challenge [1]. The potential for high-density nano-scale electronic switching using memristors [2-4] can provide a solution to scalability issues or supplement existing non-volatile random access memory applications [5, 6]. Memristors typically consist of a metal/insulator/metal stack that develops a hysteretic current-voltage loop [7, 8] after electroformation of conductive filaments by applying high voltage across insulating films. A representative listing of memristors is shown in [9]. In TaOx devices, the resistance state and switching is attributed to the concentration and motion of Oxygen vacancies in or around conductive filaments, governed by applied fields and temperature gradients [9-11]. The relatively simple construction of memristors can be straightforwardly scaled providing a path to high-density memory and potentially for high speed, low power operation, making them attractive for future memory applications. TaOx memristors have demonstrated endurance on par with state-of-the-art flash memory [12-14]. In addition, previous work has shown that TaOx memristors are radiation hard to ionizing radiation [15, 16], and displacement damage [17, 18]. The changes in the measured filament resistance induced by ion irradiations has been primarily attributed to the creation of Oxygen vacancies in the insulating film [17, 19]. However, the electro-formation process responsible for the conductive filament formation is not well controlled, remains an open topic of discussion [19, 20], and the large voltages/currents required [21] cannot be easily supplied within existing CMOS architectures. The stochastic nature of this electroforming step affects device reliability and device-to-device uniformity.

In this paper, we demonstrate a method to create electroforming-free TaOx memristors using focused ion beam irradiations. We use a direct write lithography platform to deterministically form a conductive filament in TaOx memristive devices at a targeted location. We show that this process creates fully operational memristors.

Our method uses ion irradiations to locally modify the stoichiometry of the TaOx film deterministically creating conductive filaments at specified locations. The irradiations were performed at the Ion Beam Laboratory (IBL) at Sandia National Laboratories (SNL) using the nanoImplanter (nI) which is a 100 kV focused ion beam (FIB) system (A&D

FIB100nI). The nI has a three-lens system and a Wien filter that allow for running beams from liquid metal alloy ion sources (LMAIS). For the experiments described in this paper, we used an AuSiSb alloy providing 200 keV $Si^{++}$ for irradiation with a spot size that typically is < 40 nm. The combination of a laser interferometry driven stage and direct write patterning using a Raith Elphy Plus pattern generator allows for targeted implantation with < 35 nm placement accuracy.

Electrical probes (Kleindike MM3A) enable *in-situ* electrical characterization of the devices during irradiation. The form/set-reset/loop sequences were performed using a Keithley source measure unit with the devices either under vacuum or in air. The instantaneous device resistance for a fixed voltage of -50 mV, as well as, the X and Y locations of the ion beam were recorded to spatially map out the changes in resistance.

An optical image of an SNL crossbar TaOx memristive device similar to those used in these experiments is shown in Fig. 1(a). Figure 1(b) shows the stack which consists of (from bottom to top): Si substrate, 5 nm of Ti (adhesion layer), 50 nm of Pt (bottom electrode, BE), 10 nm of TaOx, 50 nm of Ta, and 50 nm of Pt (top electrode, TE). The TaOx films were grown at SNL using ion-assisted deposition. Each die contains a series of cross-bar devices ranging in size from 500x500 $nm^2$ to 10x10 $\mu m^2$, with five structures per size. Four of these structures per size were used for ion beam irradiation while one of these structures was maintained as a control. The control was formed using conventional electroforming, ensuring that (1) electroforming was required and (2) the expected memristive behavior was observed. Typically, for unformed devices, a high resistance (>200 GΩ) is measured. In the electroforming process, we source a voltage with a current compliance set sufficiently low to avoid damaging the TaOx film. As we increase the magnitude of the sourced voltage, the device resistance decreases and we observe an increase in the measured current until the compliance is reached. Typical electroforming voltages range from 4 to 7 V with a compliance as high as 1 to 3 mA. After electroforming is completed, standard switching sweeps between the high resistance state (HRS) and the low resistance state (LRS) are accomplished by sourcing a voltage ($V_s$) and limiting the current ($I_{lim}$) on the negative side (-V and -I) of the switching sweep (or hysteretic loop) and sourcing a current ($I_s$) and limiting the voltage ($V_{lim}$) for the positive side (+V and +I) of the sweep. Typical parameters for switching are: $V_s$ = -1.5 V, $I_{lim}$ = 0.1 A, $I_s$ = 1 mA and $V_{lim}$ = 2 V. For electroformed devices, we typically see ~5 kOhm/<500 Ohm for the HRS/LRS, respectively. No difference in electrical characteristics of electroformed devices could be correlated to size and we typically observed an 80% yield of electroformed devices indicating the good quality of the TaOx film (in agreement with previous results [22]).

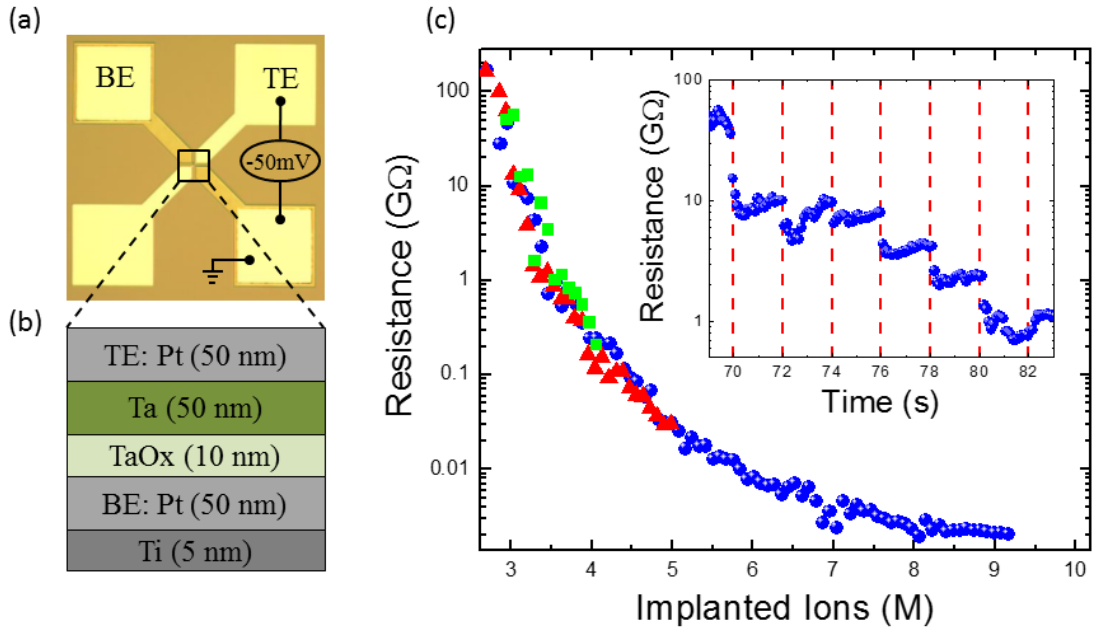

**Figure 1|** Ion beam formation of conductive filaments in TaOx memristive devices. (a) Optical view graph of a SNL crossbar memristive device showing the top electrode (TE) and bottom electrode (BE), along with the typical biasing to monitor current during irradiation. (b) Nominal memristor stack. (c) The TaOx devices typically start in the virgin state at >200 GΩ, the resistance decreases under ion irradiation until the implantation is stopped at ~4.1, 5.0, and 9.2 M ions for three example devices. Inset: Stair-step resistance decrease for the blue trace which correspond to the discrete ion implantation exposures of ~85,000 ions per step.

The resistance of these structures decreases with ion irradiation as the TaOx film becomes more conductive due to the creation of O-vacancies by the 200keV $Si^{++}$ irradiation [15, 18]. Conductive filaments in TaOx have been modeled as regions of higher concentration of O-vacancies and, therefore, a higher conductivity [11]. The ion irradiation was performed using a series of 30 ms ion pulses with a 2 s period, where each pulse consists of many ions (~100,000) targeted at the center of the crossbar structure. Figure 1(c) shows a series of resistance traces as a function of the number of ions during the irradiation for different devices illustrating the reproducibility of this technique. Figure 1(c) inset shows a stair step change in resistance due to the ion pulses for one of the irradiations (blue trace in Fig. 1(c)). This suggests that memristors are inherently radiation hard since it takes millions of ions to significantly change the resistance of a virgin film and, in agreement with our previous results, indicates that ~100 ions are needed to affect the resistance of a functioning device [19]. The insensitivity to ionizing radiation [15, 16] and displacement damage can therefore be attributed to the large number of O-vacancies needed exactly at the location of the filament to cause a change in resistance. Figure 2 illustrates our control over the final resistance of the ion beam implanted devices. We identify three regions of interest which we label high, medium and low final resistances.

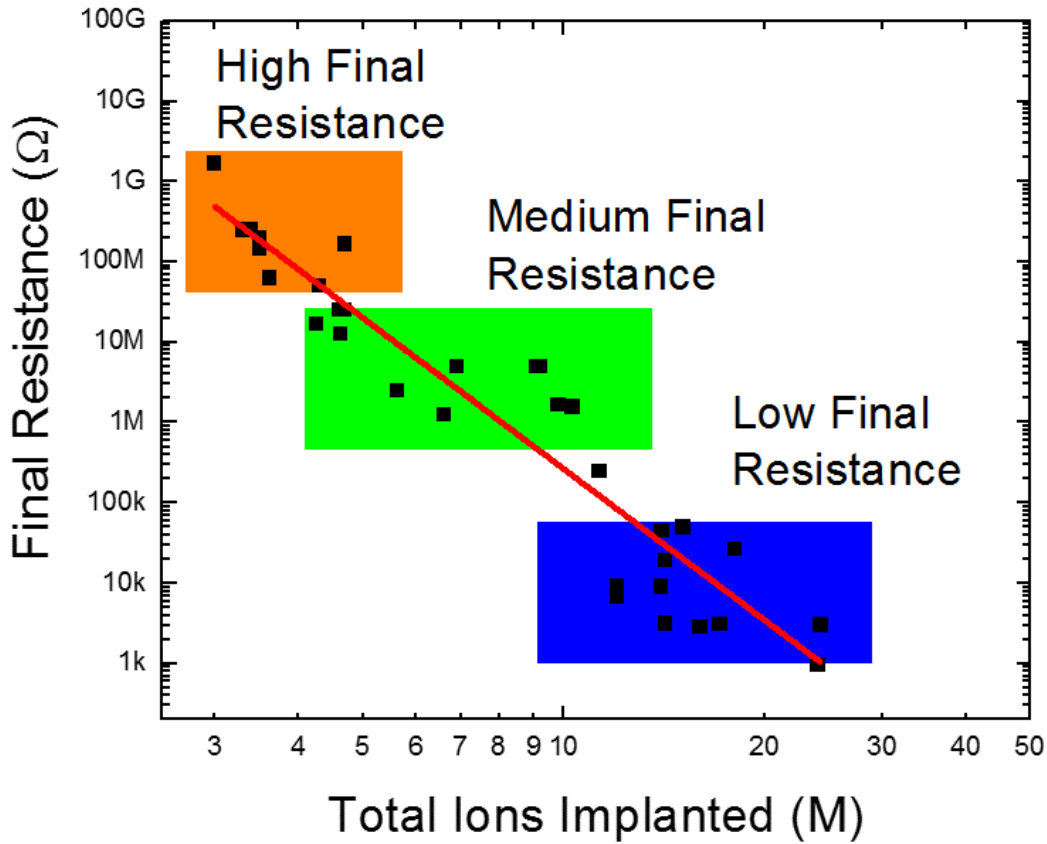

**Figure 2|** Final resistance of TaOx devices after ion beam implantation. Three regimes are qualitatively identified as high (orange), medium (green), and low (blue) final resistance.

We observe three general trends for ion irradiated devices: (1) For devices left in the high final resistance regime (≥30 MΩ, orange region) electroforming was required before switching was observed. Figure 3(a) and 3(b) show the electroforming and switching sweeps, respectively for a typical high final resistance device. (2) For devices left in the medium final resistance (~300 kΩ to ~20 MΩ, green region) the characteristic hysteretic looping was not immediate but no electroforming was necessary. (3) For devices left in the low final resistance regime (~1 kΩ to ~100 kΩ, blue region) we observed immediate hysteretic looping with no electroforming. These devices can be switched immediately after irradiation using a standard switching sweep (parameters described above). Figure 3(d) shows the resulting hysteretic looping. Electrical characterization of these devices shows that we have identified an irradiation condition that produces electroforming-free memristors that are fully operational with HRS/LRS similar to electroformed devices from the same die.

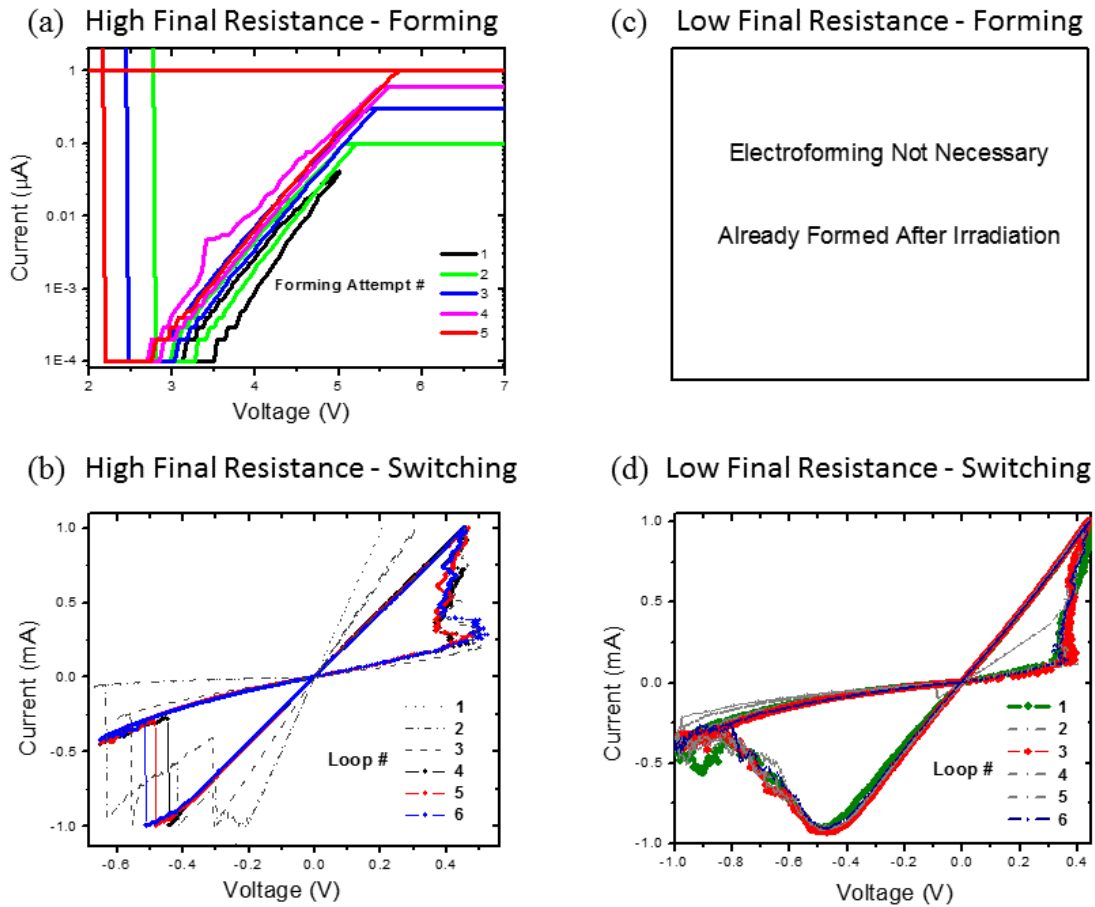

**Figure 3|** Shows plots of the electroformation (a) and hysteretic switching loops (b) for devices left in the high final resistance state (orange region of Fig. 2). This switching behavior is similar to un-irradiated electroformed devices. For devices left in the low final resistance state (blue region of Fig 2.) electroforming is not required (c) and the hysteretic switching loops (d) are immediately observed after irradiation. Not shown, devices in the medium final resistance (green region of Fig. 2) typically evolve to a fully functional memristor over a few switching loops.

Localization studies of the conductive filaments in ion beam formed devices show that these filaments are located at the targeted location. Ion beam X-Y scans with 40 nm step size, with 50-100 ions/step, to cover the cross-bar were performed on both electroformed and ion beam formed devices in the HRS state. By mapping the changes in resistance >500 $\Omega$, we can determine the location of the conductive filaments (using the same technique as in [19]). After each X-Y scan, the devices are reset to the HRS state. On devices that were electroformed, we find the resistance changes occur primarily on the edges as shown in Fig. 4(a), in agreement with our previous results [19, 20]. On multiple ion beam-formed devices we consistently find that the resistance changes are primarily found at the center of the device, as shown in Fig. 4(b), which corresponds to the targeted location. The error bars in Fig. 4 represent the uncertainty in the beam location. Fig. 4(c) shows the location of the ion beam formed conductive filament from multiple scans (resetting to the HRS state between each scan); the average shows that the most probable location is ±17 nm from center, and the standard deviation shows that the extent of the filament is ~150 nm in diameter. This agrees with other measurements of conductive filaments in TaOx memristors [24] and represents an over estimate when compared to high resolution *in-situ* filament imaging techniques [28]. This shows that we deterministically control the location of the conductive filaments.

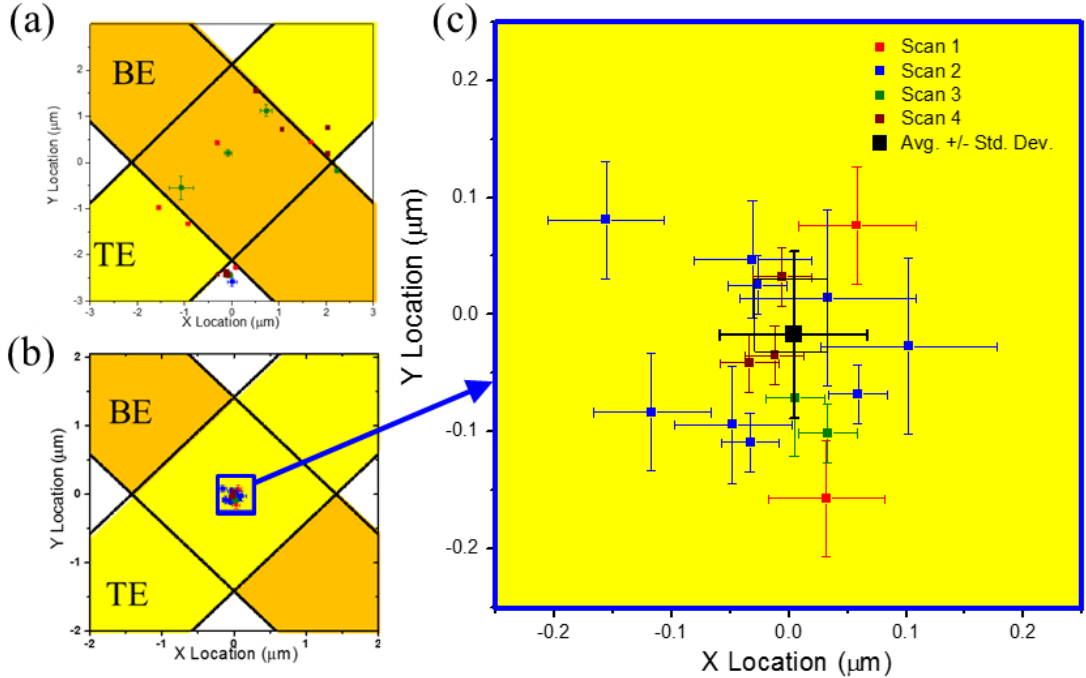

**Figure 4|** Conductive filament localization using ion beam X-Y raster scans on (a) an electroformed device where the sensitive areas are primarily localized to the edges and (b) an ion beam formed device with the sensitive area now located at the center. (c) Zoomed in view showing that the filament is at the center (±17 nm) of the device with an estimated size of ~150 nm.

The stochastic nature of TaOx film breakdown during electroforming is random and uncontrolled but this is key to enabling hysteretic switching in electroformed devices. Issues associated with electroformed devices are multiple filaments on the same device [25], device-to-device variability [26], short or open circuits, all of which are factors that directly affect device HRS/LRS ratio uniformity and yield. The conductive filament location is thought to either develop along defects or grain boundaries that act as high diffusion paths for oxygen vacancies in TiOx devices [21] or at locations where high electric fields develop due to the electrode topology. In our electroformed cross-bar structures, we observe the conductive filaments to be predominately located at the edges where we have a topology that produces a region of thinner oxide film [19]. Planar constructions to minimize edge effects exist [13] but filament formation and location are still governed by random and stochastic processes. Proposals to eliminate the electroforming by reducing the insulator film thickness, optimizing the stoichiometry [22, 27], and creation of vacancies using broad beam irradiations [18] still rely on as grown fortuitous defects.

We can avoid many of these issues by creating targeted vacancies and seeding/forming the filament at the center of the cross-bar structure where the oxide film is uniform. This ion beam irradiation approach can be used to supplement or completely avoid electroformation and enables filament formation where the TaOx film can be more uniform, potentially improving endurance and reliability. Targeting a medium final resistance state (see. Fig. 2) can potentially enable analog operation via partial filament formation. We believe that the HRS/LRS ratio, characteristic of devices tested here, are governed by film stoichiometry, device dimensions and topology, and are marginally affected by the level of irradiation as we typically see similar HRS/LRS ratios for all the ion beam formed memristive devices even at different levels of ion implantation. Future experiments can explore the optimize of the TaOx film stoichiometry and device topology and could result in HRS/LRS ratios that can be controlled by the ion irradiation.

In addition, using focused ion beam irradiations to deterministically place the filament at the targeted location enables high-density fabrication. As presented here, the device formation process is compatible with standard CMOS process flow and existing memristive devices. Full wafer fabrication using multi e-beam lithography patterning of nm-sized holes at the location where the filaments are desired could be added to standard BEOL processing steps enabling wafer scale fabrication of memristor based memories.

We have performed focused ion beam irradiations that resulted in an electroforming free memristive device where the final device resistance and the filament location is defined by ion irradiation. This process results in fully operational TaOx memristors thereby avoiding stochastic issues associated with electroformation and the incompatibility of electroforming with standard voltages and currents that drive transistor logic in CMOS circuits. Focused ion beam irradiations were used to determine the irradiation parameters that could be employed with surface masks and broad beam implantation presenting a path to CMOS compatible wafer scale fabrication of fully operational memory architectures based on memristive devices.

**Acknowledgements:**
The authors thank R. Goeke for support with device fabrication. Sandia National Laboratories is a multimission laboratory managed and operated by National Technology and Engineering Solutions of Sandia LLC, a wholly owned subsidiary of Honeywell International Inc. for the U.S. Department of Energy's National Nuclear Security Administration under contract DE-NA0003525.